\documentclass[10pt,conference]{IEEEtran}

\usepackage{graphicx}
\usepackage{subfigure}
\usepackage{cite}
\usepackage{amssymb, amsmath}

\IEEEoverridecommandlockouts

\begin{document}

\bstctlcite{TurboRef:BSTcontrol}

\title{A Union Bound Approximation for Rapid Performance Evaluation of Punctured Turbo Codes}


\author{\authorblockN{Ioannis Chatzigeorgiou, Miguel R. D. Rodrigues, Ian J. Wassell}
\authorblockA{Digital~Technology~Group,~Computer~Laboratory\\
University~of~Cambridge,~United~Kingdom\\
Email:~\{ic231,~mrdr3,~ijw24\}@cam.ac.uk} \and
\authorblockN{Rolando Carrasco}
\authorblockA{School~of~EE\&C~Engineering\\
University~of~Newcastle,~United~Kingdom\\
Email:~r.carrasco@ncl.ac.uk}
\thanks{This work is supported by EPSRC
under Grant GR/S46437/01.}}

\maketitle

\begin{abstract}
In this paper, we present a simple technique to approximate the
performance union bound of a punctured turbo code. The bound
approximation exploits only those terms of the transfer function
that have a major impact on the overall performance. We revisit the
structure of the constituent convolutional encoder and we develop a
rapid method to calculate the most significant terms of the transfer
function of a turbo encoder. We demonstrate that, for a large
interleaver size, this approximation is very accurate. Furthermore,
we apply our proposed method to a family of punctured turbo codes,
which we call pseudo-randomly punctured codes. We conclude by
emphasizing the benefits of our approach compared to those employed
previously. We also highlight the advantages of pseudo-random
puncturing over other puncturing schemes.
\end{abstract}

\IEEEpeerreviewmaketitle

\section{Introduction}
\label{Intro}

Turbo codes, originally conceived by Berrou \textit{et al}.
\cite{Berrou96} are widely known for their astonishing performance
on the additive white Gaussian noise (AWGN) channel. Methods to
evaluate an upper bound on the bit error probability (BEP) of a
parallel-concatenated coding scheme have been proposed by Divsalar
\textit{et al}. \cite{Divsalar95} as well as Benedetto and Montorsi
\cite{Benedetto96a}. In addition, guidelines for the optimal design
of the constituent convolutional codes were presented in
\cite{Benedetto96b}.

The rate of a turbo code can be increased by puncturing the outputs
of the turbo encoder. Guidelines and design considerations for
punctured turbo codes have been derived by analytical
\cite{Acikel99,Babich02,Kousa02} as well as simulation-based
approaches \cite{FanMo99,Land00}. Upper bounds on the bit error
probability (BEP) can be easily evaluated based on the techniques
presented in \cite{Kousa02} and \cite{Chatzigeorgiou06b}. However,
computation of the upper bound can be complex and time-consuming,
when a large interleaver size and certain puncturing patterns are
used.


The motivation for this paper is to derive simple expressions for
the calculation of the dominant term of the performance union bound
for punctured parallel concatenated convolutional codes (PCCCs).
Previously, complex approaches based on the full transfer function
of each constituent code, have been used. In Section
\ref{PerfEvalPunc} we demonstrate that for a large interleaver size,
the dominant term can be used as an accurate approximation of the
overall performance union bound. In Section \ref{Structure_PuncRSC}
we analyze the properties of constituent convolutional encoders so
as to obtain exact expressions for the dominant term. A case study
considering pseudo-random puncturing is presented in Section
\ref{CaseStudy} and the paper concludes with a summary of the main
contributions.


\section{An Upper Bound to the Error Probability of Punctured Turbo Codes and its Approximation}
\label{PerfEvalPunc}

Turbo codes, in the form of rate-1/3 PCCCs, consist of two rate-1/2
recursive systematic convolutional (RSC) encoders separated by an
interleaver of size $N$ \cite{Berrou96}. The information bits are
input to the first constituent RSC encoder, while an interleaved
version of the information bits are input to the second RSC encoder.
The output of the turbo encoder consists of the systematic bits of
the first encoder, which are identical to the information bits, the
parity-check bits of the first encoder and the parity-check bits of
the second encoder.

Rates higher than 1/3 can be obtained by periodic elimination of
specific codeword bits from the output of a rate-1/3 turbo encoder.
Punctured codes are classified as systematic (S), partially
systematic (PS) or non-systematic (NS) depending on whether all,
some or none of their systematic bits are transmitted \cite{Land00}.
Note that a punctured PCCC can also be seen as a PCCC constructed
using two constituent punctured RSC codes.

Puncturing of a rate-1/2 RSC to obtain a higher rate RSC is
represented by an $2\times M$ matrix as follows:
\begin{equation}
\label{punc_pattern} \mathbf{P}=\left[
\begin{matrix}
\mathbf{P}_{U}\\
\mathbf{P}_{Z}
\end{matrix}\right]=\left[
\begin{matrix}
p_{1,1}&p_{1,2}&\ldots&p_{1,M}\\
p_{2,1}&p_{2,2}&\ldots&p_{2,M}
\end{matrix}\right],
\end{equation}where $M$ is the puncturing period and $p_{i,m}\in\{0,1\}$, with $i\!=\!1,2$ and
$m\!=\!1,\ldots,M$. For $p_{i,m}\!=\!0$, the corresponding output
bit is punctured. The puncturing pattern $\mathbf{P}$ for the
rate-1/2 encoder consists of the puncturing vector $\mathbf{P}_{U}$
for the systematic output sequence and the puncturing vector
$\mathbf{P}_{Z}$ for the parity-check output sequence.

It was shown in \cite{Divsalar95} and \cite{Benedetto96a} that
performance bounds for a PCCC can be obtained from the transfer
functions, or equivalently the weight enumerating functions (WEFs),
of the terminated constituent RSC codes. A WEF provides all paths of
length $N$ that start from the zero state, can remerge with and
diverge from the zero state more than once, and terminate at the
zero state.

More specifically, the conditional WEF (CWEF) of a punctured
convolutional code $\mathcal{C}'$, denoted as
$A^{\mathcal{C}'}(w,U,Z)$, assumes the form \cite{Benedetto96a}
\begin{equation}
\label{Cprime_CWEF}
\small{A^{\mathcal{C}'}(w,U,Z)=\:\sum\limits_{u}\sum\limits_{z}
A^{\mathcal{C}'}_{w,u,z}U^{u}Z^{z}},
\end{equation}where $A^{\mathcal{C}'}_{w,u,z}$ is the number of codeword sequences composed of a
systematic and a parity-check sequence having weights $u$ and $z$,
respectively, which were generated by input sequences of a given
weight $w$. The overall weight of a codeword sequence is $u+z$.

The input-redundancy WEF (IRWEF), $A^{\mathcal{C}'}(W,U,Z)$,
provides all codeword sequences for all possible values of input
information weight, and is related to the CWEF as follows
\cite{Benedetto96a}
\begin{equation}
\label{Cprime_CWEFtoIRWEF}
\small{A^{\mathcal{C}'}(W,U,Z)=\:\sum\limits_{w}
A^{\mathcal{C}'}(w,U,Z)W^{w}}.
\end{equation}

A relationship between the CWEF of a PCCC and the CWEFs of the
constituent codes, $\mathcal{C}'_{1}$ and $\mathcal{C}'_{2}$
respectively, can be easily derived only if we assume the use of a
uniform interleaver of size $N$, an abstract probabilistic concept
introduced in \cite{Benedetto96a}. More specifically, if
$A^{\mathcal{C}'_{1}}(w,U,Z)$ and $A^{\mathcal{C}'_{2}}(w,U,Z)$ are
the CWEFs of the constituent codes, the CWEF of the PCCC,
$A(w,U,Z)$, is equal to
\begin{equation}
\label{PCCCprime_CWEF} \small{A(w,U,Z) =
\frac{A^{\mathcal{C}'_{1}}(w,U,Z)\cdot
A^{\mathcal{C}'_{2}}(w,U=1,Z)} {\displaystyle \binom{N}{w}}}.
\end{equation}The systematic output sequence
of the second constituent encoder is not transmitted, therefore it
does not contribute to the overall weight of the turbo codeword
sequences, so it is eliminated by setting $U\!\!=\!\!1$ in
$A^{\mathcal{C}'_{2}}(w,U,Z)$. The IRWEF of the PCCC, $A(W,U,Z)$,
can be computed from the CWEF, $A(w,U,Z)$, in a manner identical to
(\ref{Cprime_CWEFtoIRWEF}).

The input-output weight enumerating function (IOWEF) provides the
number of codewords generated by an input sequence of information
weight $w$, whose overall weight is $d$, in contrast with the IRWEF,
which distinguishes between the systematic and the parity-check
weights. For the case of a punctured PCCC, the corresponding IOWEF
assumes the form
\begin{equation}
\label{PCCCprime_IOWEF}
\small{B(W,D)=\:\sum\limits_{w}\sum\limits_{d}B_{w,d}W^{w}D^{d}},
\end{equation} where the coefficients $B_{w,d}$ can be derived
from the coefficients $A_{w,u,z}$ of the IRWEF, based on the
expression
\begin{equation}
\label{PCCCprime_IOWEF_COEFF}
\small{B_{w,d}=\!\sum\limits_{u+z=d}A_{w,u,z}}.
\end{equation}

The IOWEF coefficients $B_{w,d}$ can be used to determine a tight
upper bound, denoted as $P^{u}_{b}$, on the BEP $P_{b}$, for
maximum-likelihood (ML) soft decoding for the case of an AWGN
channel, as follows \cite{Benedetto96a}
\begin{equation}
\label{PB}\small{P_{b}\leq P^{u}_{b}=\sum\limits_{w}P(w)},
\end{equation}where $P(w)$ is the union bound of all error
events with information weight $w$, and is defined as
\begin{equation}
\label{Pw}\small{P(w)=\:\sum\limits_{d}\frac{w}{N}B_{w,d}
Q\left(\sqrt{\frac{2R\cdot E_{b}}{N_{0}}\cdot d} \right)},
\end{equation}where $R$ is the rate of the punctured turbo code.

\begin{figure}[t]
    \centering
    \includegraphics[width=0.80\linewidth]{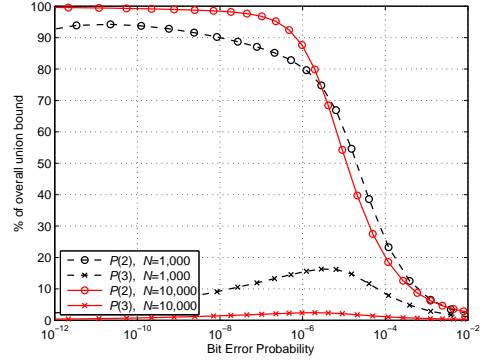}
    \caption{Contribution to the union bound for rate-1/2 S-PCCC(1,17/15,17/15) using an interleaver of size $N\!=\!1,000$ and $N\!=\!10,000$ bits}
    \label{figContrib}
\end{figure}

In \cite{Benedetto96b}, Benedetto \textit{et al.} investigated the
performance of rate-1/3 PCCCs and observed that the union bound
$P(w_{\text{min}})$ of all error events with the lowest information
weight $w_{\text{min}}$, becomes dominant as the interleaver size
$N$ increases. Owing to the structure of an RSC encoder, the minimum
information weight of a terminated RSC code is always equal to two,
i.e., $w_{\text{min}}\!=\!2$. Consequently, the overall performance
bound $P^{u}_{b}$ can be approximated by $P(2)$, when a large
interleaver size is used. The same trend is also observed in the
case of punctured turbo codes. The contribution, as a percentage, of
$P(2)$ and $P(3)$ to $P^{u}_{b}$ is illustrated in
Fig.\ref{figContrib}. As an example, rate-1/2 S-PCCC(1, 17/15,
17/15) is considered, using a uniform interleaver of size either
$N\!=\!1,000$ or $N\!=\!10,000$. It is apparent that $P(2)$ becomes
the dominant contribution over a broad range of BEP values, as the
interleaver size increases.

We see from (\ref{Pw}) that $P(2)$ depends heavily on the minimum
weight of the turbo codeword sequences, commonly known as free
effective distance $d_{\text{free,eff}}$ \cite{Benedetto96b}. We use
the notation $d^{\mathcal{C}'_{1}}_{\text{min}}$ to denote the
minimum weight of the codeword sequences generated by the first
constituent encoder and $z^{\mathcal{C}'_{2}}_{\text{min}}$ to
denote the minimum weight of the parity-check output sequences,
generated by the second constituent encoder. In both cases an input
sequence of information weight 2 is assumed. Therefore, the free
effective distance of a PCCC can be expressed as
\begin{equation}
\label{free_effective_generic}
d_{\text{free,eff}}=d^{\mathcal{C}'_{1}}_{\text{min}}+z^{\mathcal{C}'_{2}}_{\text{min}}.
\end{equation}The free effective distance is the most significant
parameter that influences the PCCC performance. The constituent
encoders should be chosen to maximise
$d^{\mathcal{C}'_{1}}_{\text{min}}$ and
$z^{\mathcal{C}'_{2}}_{\text{min}}$, and consequently
$d_{\text{free,eff}}$.


\section{Computing the Upper Bound Approximation}
\label{Structure_PuncRSC}

In order to compute $P(2)$ and thus obtain a good approximation to
the overall performance bound $P^{u}_{b}$, we only need to calculate
the CWEF of each constituent code for $w\!=\!2$, i.e.,
$A^{\mathcal{C}'_{1}}(2,U,Z)$ and $A^{\mathcal{C}'_{2}}(2,U,Z)$.

Both CWEFs could be obtained by brute-force, i.e., input all
possible sequences of weight 2 to each constituent encoder and group
the output codeword sequences according to their systematic and
parity-check weights. Although this approach is conceptually simple,
it is extremely time-consuming, especially when a large interleaver
size is used.

The techniques proposed in \cite{Kousa02} and
\cite{Chatzigeorgiou06b} are more complex but less time-consuming.
They both use the state diagram of a parent RSC code and introduce
the puncturing patterns to obtain the full CWEF of the corresponding
punctured RSC code. However, for large interleaver sizes and
puncturing patterns with a long period, complexity becomes a
prohibiting factor for the implementation of either approach.

In this section we use the properties of the trellis structure of
RSC codes to express the CWEF, for $w\!=\!2$, of an RSC encoder as a
function of its memory size, generator polynomials, and puncturing
pattern. Consequently, derivation of the state equations and
computation of the full transfer function of each constituent code,
required in \cite{Kousa02} and \cite{Chatzigeorgiou06b}, is not
necessary. Hence, PCCCs using both a large interleaver and a long
puncturing pattern can now be easily supported.

\subsection{Unpunctured Rate-1/2 RSC Encoders}
A rate-1/2 RSC encoder, $\mathcal{C}$, is characterised by its
feedback and feedforward polynomials, $G_{R}(D)$ and $G_{F}(D)$
respectively. The degree of each polynomial is equal to the memory
size $\nu$ of the encoder. A hypothesis commonly made
\cite{Berrou96,Benedetto96b} so as to facilitate analysis of RSC
codes is that $G_{F}(D)$ is a monic function and that the initial
state of the encoder is the zero state, for every input sequence of
length $N$.

\begin{figure}[b]
    \centering
    \includegraphics[width=0.85\linewidth]{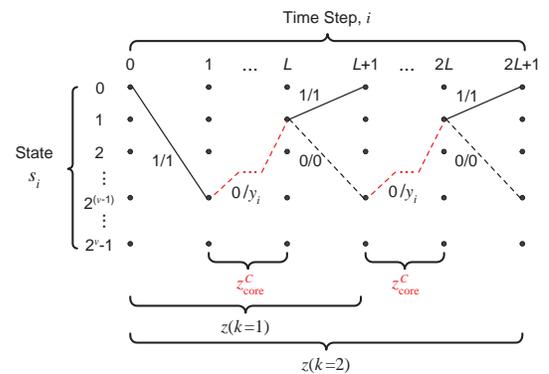}
    \caption{Trellis diagram for codeword sequences of information weight
    two. Dashed lines correspond to paths generated when the
    rate-1/2 RSC encoder operates as a pseudo-random generator. A pair $x_{i}/y_{i}$ next to a branch denotes the input and
    parity-check output bits, respectively, generated at the end of a state transition.}
    \label{RSC_Trellis_Short}
\end{figure}

Input sequences of weight 2 force the trellis path to diverge from
the zero state and re-merge with it, after a number of time-steps.
More specifically, the input sequence will change the state from $0$
to $2^{\nu-1}$, when the first non-zero bit is input to the encoder,
as it is illustrated in Fig.\ref{RSC_Trellis_Short}. For as long as
a trail of zeros follows the first non-zero input bit, the RSC
encoder behaves like a pseudo-random generator, with the same state
transitions being repeated every $L$ time-steps, where $L$ is the
period of the feedback polynomial. In order for the path to re-merge
with the zero state, the second non-zero bit should be input to the
encoder when state $1$ is reached, i.e., after $kL+1$ time-steps,
where $k\!=\!1,2,\ldots,\lfloor (N\!-\!1)/L\rfloor$ and $\lfloor
(N-1)/L\rfloor$ is the integer part of $(N\!-\!1)/L$. Furthermore,
as it is depicted in Fig.\ref{RSC_Trellis_Short}, when a non-zero
input bit causes the path to diverge from or re-merge to the zero
state, both the systematic and the parity-check outputs give a
logical 1. Therefore, if $z^{\mathcal{C}}_{\text{core}}$ is the
parity-check weight due to the transitions of the encoder from state
$2^{\nu-1}$ to state $1$, the overall weight $z$ of a parity-check
sequence can be expressed as
\begin{equation}
\label{zk} z(k)=kz^{\mathcal{C}}_{\text{core}}+2,\quad\text{for
}k\!=\!1,2,\ldots,\lfloor (N-1)/L\rfloor.
\end{equation}Note that the state sequence during the transitions
from state $2^{\nu-1}$ to state $1$ and, consequently, the value of
$z^{\mathcal{C}}_{\text{core}}$, depend on the selected feedback
polynomial. The minimum parity-check weight
$z^{\mathcal{C}}_{\text{min}}$ can be derived from $z(k)$ by setting
$k\!=\!1$, i.e.,
\begin{equation}
\label{zm_func_of_zk} z^{\mathcal{C}}_{\text{min}}=z(1).
\end{equation}

Based on (\ref{Cprime_CWEF}) and (\ref{zk}), the CWEF for $w\!=\!2$,
$A^{\mathcal{C}}(2,U,Z)$, of the rate-1/2 RSC code when no
puncturing is applied, assumes the form
\begin{equation}
\label{CWEF_B} A^{\mathcal{C}}(2,U,Z)=\:\sum\limits_{k=1}^{\lfloor
(N-1)/L\rfloor}A^{\mathcal{C}}_{2,2,z(k)}U^{2}Z^{z(k)},
\end{equation}where $A^{\mathcal{C}}_{2,2,z(k)}$ is the number of codeword sequences with parity-check weight $z(k)$, given by
\begin{equation}
\label{alpha} A^{\mathcal{C}}_{2,2,z(k)}=N-kL.
\end{equation}

When the feedback polynomial, $G_{R}(D)$, of an RSC encoder is
selected to be primitive, the encoder visits all possible
$2^{\nu}\!-\!1$ states with a maximum period of
$L\!=\!2^{\nu}\!-\!1$ time-steps \cite{Macwilliams76}, if the
information weight of the input sequence is 2. As pointed out in
\cite{Perez96}, maximization of $L$ increases the length of the
shortest weight 2 input sequence, therefore increasing the chance of
achieving a high weight $z^{\mathcal{C}}_{\text{core}}$ and,
consequently, $z^{\mathcal{C}}_{\text{min}}$. An exact expression
for $z^{\mathcal{C}}_{\text{core}}$ can be derived based on the
properties of pseudo-random sequences \cite{Macwilliams76} or the
analysis in \cite{Benedetto96b}, i.e.,
\begin{equation}
\label{zcore_primitive} z^{\mathcal{C}}_{\text{core}}=2^{\nu-1},
\end{equation}provided that $G_{R}(D)\neq G_{F}(D)$.

Since $z^{\mathcal{C}}_{\text{core}}$ only depends on the memory
size of the encoder, so does the CWEF of each constituent code,
$A^{\mathcal{C}}(2,U,Z)$ and, consequently, the union bound of all
error events with information weight $2$, $P(2)$. Therefore, the
performance of a rate-1/3 PCCC, using a large interleaver size,
mainly depends on the memory size of each constituent RSC encoder
and not the underlying code, provided that the feedforward
polynomial of each RSC encoder is different from the feedback
primitive polynomial.


\subsection{Punctured RSC Encoders}

Rates higher than 1/2 can be achieved using a $2\times M$ puncturing
pattern $\mathbf{P}$ on a parent rate-1/2 RSC encoder $\mathcal{C}$.
At a time step $i$ $(0\!\leq\!i\!<\!N)$, the weights of the
systematic and parity-check output bits of the punctured encoder
$\mathcal{C}'$ will be $x_{i}\cdot p_{1,m}$ and $y_{i}\cdot
p_{2,m}$, respectively, where $x_{i}$, $y_{i}$ are the output bits
of the parent rate-1/2 encoder and $p_{1,m}$, $p_{2,m}$ are the
elements of column-$m$ $(1\!\leq\!m\!\leq\!M)$ of the puncturing
pattern $\mathbf{P}$. Note that, owing to the systematic nature of
the encoder, $x_{i}$ also represents the input information bit. The
relationship between $m$ and $i$ is
\begin{equation}
\label{time_and_pattern} m=\text{rem}(i+1,M),
\end{equation}
where $\text{rem}(i+1,M)$ denotes the remainder from the division
$(i+1)/M$. Since the period of $\mathbf{P}$ is $M$, its elements are
repeated in such a way that $p_{1,m}\!=\!p_{1,(m+jM)}$ and
$p_{2,m}\!=\!p_{2,(m+jM)}$, where $j$ is a non-negative integer.

\begin{figure}[t]
    \centering
    \includegraphics[width=0.70\linewidth]{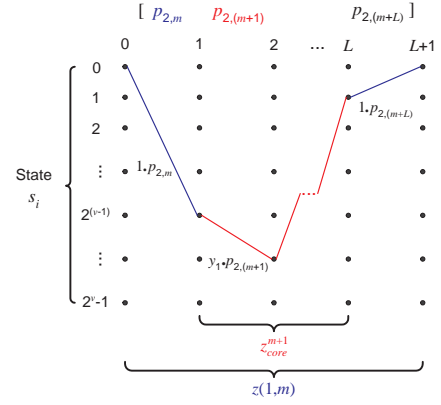}
    \caption{Trellis diagram for the weight calculation of parity-check sequences}
    \label{RSC_Trellis_Punc}
\end{figure}

In order to compute the CWEF of the punctured RSC for information
weight $w\!=\!2$, i.e., $A^{\mathcal{C}'}(2,U,Z)$, we need to derive
an expression for the weight of the systematic and parity-check
output sequences. Although information sequences with $w\!=\!2$
generate paths of length $kL\!+\!1$, we first consider paths of
length $L\!+\!1$, i.e., $k\!=\!1$, for simplicity. The weight
$u(k\!=\!1,m)$ of a systematic sequence, whose path diverges from
the zero state when $p_{1,m}$ is active, is given by
\begin{equation}
\label{u_k1} u(k\!=\!1,m)=p_{1,m}+p_{1,(m+L)},
\end{equation}
since the two non-zero bits occur at the very beginning and at the
very end of the path. Similarly, the weight $z(k\!=\!1,m)$ of the
parity-check sequence, whose path diverges from the zero state when
$p_{2,m}$ is active, assumes the form
\begin{equation}
\label{z_k1} z(k\!=\!1,m)=p_{2,m}+z^{m+1}_{\text{core}}+p_{2,(m+L)},
\end{equation}since the parity-check bits at the beginning and at the
end of the path are non-zero, while the weight of the remaining path
is $z^{m+1}_{\text{core}}$, as it is illustrated in
Fig.\ref{RSC_Trellis_Punc}. In order to calculate $z(k\!=\!1,m)$ for
every value of $m$, we first need to derive $z^{1}_{\text{core}},
z^{2}_{\text{core}}, \ldots, z^{M}_{\text{core}}$, by applying the
$M$ circularly shifted versions of the puncturing vector
$[p_{2,1},\ldots,p_{2,M}]$ to the corresponding output parity-check
bits of the parent rate-1/2 RSC encoder, i.e,
\begin{equation}
\label{z_core_punc}
z^{m}_{\text{core}}=\:\sum\limits_{i=1}^{L-1}\left(y_{i}\cdot
p_{2,(i+m-1)}\right).
\end{equation}

If we extend our analysis to codewords associated with paths of
length $kL\!+\!1$, we obtain the generic expressions for $u(k,m)$
and $z(k,m)$ as follows
\begin{equation}
\label{u_k_any} u(k,m)=p_{1,m}+p_{1,(m+kL)}
\end{equation}
\begin{equation}
\label{z_k_any}
z(k,m)=p_{2,m}+{\sum\limits_{j=0}^{k-1}z^{m+jL+1}_{\text{core}}}+
p_{2,(m+kL)},
\end{equation}
where $z^{m+jM}_{\text{core}}\!=\!z^{m}_{\text{core}}$, due to the
periodicity of the puncturing pattern. Since any codeword sequence,
generated by an input sequence of weight $2$, can be described by a
polynomial $U^{u(k,m)}Z^{z(k,m)}$ for a given $k$ and $m$, the
summation of all polynomials of the form $U^{u(k,m)}Z^{z(k,m)}$ over
all possible values of $k$ and $m$ will give the CWEF,
$A^{\mathcal{C}'}(2,U,Z)$, of the punctured RSC code
\begin{equation}
\label{CWEF_Punc}
A^{\mathcal{C}'}(2,U,Z)=\:\sum\limits_{k=1}^{\lfloor
(N-1)/L\rfloor}\sum\limits_{m=1}^{M}A^{\mathcal{C}'}_{k,m}U^{u(k,m)}Z^{z(k,m)},
\end{equation}
where $A^{\mathcal{C}'}_{k,m}$ is the total number of codeword
sequences with systematic weight $u(k,m)$ and parity-check weight
$z(k,m)$. Coefficients $A^{\mathcal{C}'}_{k,m}$ can be easily
derived if we observe that there are $N-kL$ codeword sequences of
length $kL+1$ each. The codeword sequences are grouped into $M$
groups, whose members share the same weights $u(k,m)$ and $z(k,m)$.
Thus, the number of codeword sequences in the $m$-th group is given
by
\begin{equation} \label{alpha_Punc}
{\small A^{\mathcal{C}'}_{k,m}=\left\{\begin{array}[\relax]{ll}
\left\lfloor\frac{N-kL}{M}\right\rfloor,& \; \text{if rem}\left((N-kL),M\right)<m\\[3pt]
\left\lfloor\frac{N-kL}{M}\right\rfloor+1,& \; \text{otherwise.}
\end{array}\right.}
\end{equation}

\begin{figure}[t]
    \centering
    \includegraphics[width=0.80\linewidth]{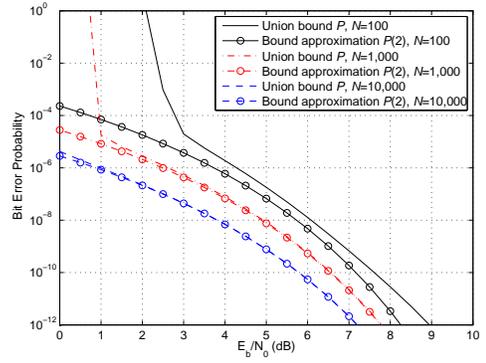}
    \caption{Exact union bounds and their approximation for a rate-1/2 S-PCCC(1,17/15,17/15) using interleavers of various sizes}
    \label{figBoundsRate12}
\end{figure}

Using (\ref{CWEF_Punc}), we can accurately and efficiently derive
$P(2)$, i.e., the probability of all error events with information
weight 2, which is a good approximation of the union bound
$P^{u}_{b}$, for a large interleaver size. In the example shown in
Fig.\ref{figBoundsRate12}, we see that $P(2)$ closely matches
$P^{u}_{b}$, when the interleaver reaches the size of $N\!=\!10,000$
bits.

\section{Case Study: Pseudo-random Puncturing}
\label{CaseStudy}

In this section we consider constituent RSC encoders employing
primitive feedback polynomials, therefore the period $L$ assumes the
maximum value of $2^{\nu}-1$. Furthermore, we assume that the
elements of the puncturing vector $\mathbf{P}_{Z}$, for the
parity-check output, form a pseudo-random sequence of period
$M\!=\!L$, generated by the same primitive polynomial as that of the
RSC encoder.

Since the puncturing period $M$ is equal to the period $L$ of the
feedback polynomial, $u(k,m)$ and $z(k,m)$ are reduced to
\begin{equation}
\label{u_k_any_sp1} u(k,m)=u(m)=2p_{1,m},
\end{equation}
\begin{equation}
\label{z_k_any_sp1} z(k,m)=kz^{m+1}_{\text{core}}+2p_{2,m}.
\end{equation} Calculation of $z(k,m)$ and, consequently,
$A^{\mathcal{C}'}(2,U,Z)$, requires knowledge of the $L$ values of
$z^{m}_{\text{core}}$. However, the assumption of pseudo-random
puncturing can further simplify the computation of $z(k,m)$.


\subsection{Derivation of the Minimum Weight Values}
\label{Configurations_Punc}

In order to express $z^{m}_{\text{core}}$ in a more compact form, we
first need to consider the autocorrelation function $\phi(j)$ of a
polar sequence, which is defined as \cite{Proakis01}
\begin{equation}
\label{autocorrelation_polar}
\phi(j)=\:\sum\limits_{i=1}^{L}(2y_{i}-1)(2y_{i+j}-1)
\end{equation}
where $y_{i}\!=\!\{0,1\}$ is the output of the parent rate-1/2 RSC
encoder at time-step $i$ for an input sequence of information weight
2, and $0\leq j<L$. The parity-check sequence generated during the
time period from $i\!=\!1$ until $i\!=\!L$ is a pseudo-random
sequence, provided that the encoder does not return to the zero
state. In this case, the autocorrelation function can be reduced to
\cite{Macwilliams76,Sarwate80}
\begin{equation}
\label{autocorrelation_polar_PRBS}
\phi(j)=\left\{\begin{array}[\relax]{ll}
2^{\nu}-1,& \quad \text{if }j=0\\
-1,& \quad \text{if }1\leq j<L.
\end{array}\right.
\end{equation}Combining (\ref{autocorrelation_polar}) and
(\ref{autocorrelation_polar_PRBS}) we find that
\begin{equation}
\label{autocorrelation_AND} \sum\limits_{i=1}^{L}(y_{i}\cdot
y_{i+j})=\left\{\begin{array}[\relax]{ll}
2^{\nu-1},& \quad \text{if }j=0\\
2^{\nu-2},& \quad \text{if }1\leq j<L.
\end{array}\right.
\end{equation}Since the puncturing vector for the parity-check bits, $\mathbf{P}_{Z}\!=\![p_{2,i}]$, is
also a pseudo-random sequence generated by the same primitive
polynomial $G_{R}(D)$, such that $p_{2,i+1}\!=\!y_{i}$, we can
rewrite expression (\ref{autocorrelation_AND}) as follows
\begin{equation}
\label{autocorrelation_punc} \sum\limits_{i=1}^{L}(y_{i}\cdot
p_{2,(i+m)})=\left\{\begin{array}[\relax]{ll}
2^{\nu-1},& \quad \text{if }m=1\\
2^{\nu-2},& \quad \text{if }2\leq m\leq L
\end{array}\right.
\end{equation}where $j$ was replaced by $m\!-\!1$. Due to the structure of the RSC encoder, the last bit of the parity-check sequence is always zero, i.e., $y_{L}\!=\!0$, therefore
\begin{equation}
\label{z_core_sp1} \sum\limits_{i=1}^{L-1}(y_{i}\cdot
p_{2,(i+m)})=z^{m+1}_{\text{core}}=\left\{\small{\begin{array}[\relax]{ll}
2^{\nu-1},& \text{if }m=1\\
2^{\nu-2},& \text{if }2\leq m<L
\end{array}}\right.
\end{equation}so $z^{m}_{\text{core}}$ is now a function of the memory $\nu$ of the RSC
encoder. Having in mind that $p_{2,L+1}\!=\!0$ since $y_{L}\!=\!0$,
and so is $p_{2,1}$, and that $L\!=\!2^{\nu}-1$, the weight of a
parity-check sequence assumes the form
\begin{equation}
\label{z_k_any_sp2} z(k,m)=\left\{\begin{array}[\relax]{ll}
k2^{\nu-1},& \quad \text{if }m\!=\!1\\
k2^{\nu-2}+2p_{2,m},& \quad \text{if }2\leq m\leq2^{\nu}\!-1
\end{array}\right.
\end{equation}where $2^{\nu-1}$ elements of the puncturing vector $\mathbf{P}_{Z}$
are equal to 1 and the remaining $2^{\nu-1}\!-\!1$ are equal to 0,
since the elements of $\mathbf{P}_{Z}$ form a pseudo-random sequence
\cite{Macwilliams76,Sarwate80}.

The minimum weight of the parity-check sequences,
$z^{\mathcal{C}'}_{\text{min}}$ can be expressed as
\begin{equation}
\label{z_min_punc} z^{\mathcal{C}'}_{\text{min}}=\min_{m=1\ldots
L}\left\{z(k=1,m)\right\}=\left\{\begin{array}[\relax]{ll}
2,& \text{for }\nu=2\\
2^{\nu-2},& \text{for }\nu>2
\end{array}\right.
\end{equation}whereas the minimum weight of the codeword sequences,
$d^{\mathcal{C}'}_{\text{min}}$ assumes the form
\begin{equation}
\label{d_min_punc} d^{\mathcal{C}'}_{\text{min}}=\min_{m=1\ldots
L}\left\{u(m)+z(k=1,m)\right\}.
\end{equation}

As in the case of rate-1/3 PCCCs, we conclude that when a large
interleaver is used, the performance of a PCCC whose parity-check
sequences were punctured using pseudo-random patterns, mainly
depends on the memory size of the constituent RSC encoders, and not
the exact underlying codes.


\subsection{Example Configurations for Rate-1/2 PCCCs}
\label{Configurations_Punc}

In order to maximize the minimum weight of the codeword sequences,
$d^{\mathcal{C}'_{1}}_{\text{min}}$, generated by the first
constituent RSC encoder of a PCCC, we can set the puncturing vector
for the systematic output, $\mathbf{P}_{U}\!=\![p_{1,m}]$, to be the
complement of the puncturing vector for the parity-check output
$\mathbf{P}_{Z}\!=\![p_{2,m}]$, i.e.,
$p_{1,m}\!=\!\overline{p_{2,m}}$. This configuration prevents $u(m)$
and $z(k,m)$ from assuming the smallest values at the same time.
Therefore, expression (\ref{d_min_punc}) becomes
\begin{equation}
\label{d_min_PS-RSC} d^{\mathcal{C}'_{1}}_{\text{min}}=2+2^{\nu-2},
\end{equation}for $\nu\geq2$.

A code rate of 1/2 can be achieved, if the parity-check output of
the second RSC encoder is not punctured. In that case,
$z^{\mathcal{C}'_{2}}_{\text{min}}$ can be derived from
(\ref{zm_func_of_zk}) and (\ref{zcore_primitive}). The free
effective distance of the corresponding PS-PCCC assumes the form
\begin{equation}
\label{dfree_rate12_PseudoA} d_{\text{free,eff}}=4+3(2^{\nu-2}).
\end{equation}We refer to this example configuration as ``Pseudo A''.

If our objective is to obtain a turbo code whose BEP performance
quickly converges to the union bound but experiences a high error
floor, we need to increase the number of transmitted systematic bits
\cite{Crozier05,Chatzigeorgiou06b,Blazek02}. The parity-check output
of both the first and the second constituent encoder is punctured
using the same vector $\mathbf{P}_{Z}$. Bearing in mind that
$\mathbf{P}_{U}$ is taken to be the complement of $\mathbf{P}_{Z}$,
we need to replace all but one of the 0's in $\mathbf{P}_{U}$ with
1's, in order to achieve a code rate of 1/2. The minimum codeword
weight $d^{\mathcal{C}'_{1}}_{\text{min}}$ for the first constituent
encoder is given by (\ref{d_min_PS-RSC}), while the minimum
parity-check weight $z^{\mathcal{C}'_{2}}_{\text{min}}$ for the
second constituent encoder is given by (\ref{z_min_punc}). The
summation of the two minimum weights yields the free effective
distance of the PS-PCCC
\begin{equation}
\label{dfree_rate12_PseudoB}
d_{\text{free,eff}}=\left\{\begin{array}[\relax]{ll}
5,& \quad \text{for }\nu=2\\
2+2^{\nu-1},& \quad \text{for }\nu>2.
\end{array}\right.
\end{equation}We refer to this example configuration as ``Pseudo B''.

The particular puncturing patterns of each example configuration for
the case of PCCC(1, 17/15, 17/15) are presented in Table
\ref{PuncPatterns}. The configuration denoted as ``Litt A'' achieves
a very low error floor and it was obtained through exhaustive search
using \cite{Chatzigeorgiou06b}, whereas ``Litt B'' is the
conventional approach for obtaining rate-1/2 turbo codes.


\subsection{The Benefits of Pseudo-random Patterns} \label{benefits}

Good punctured PCCCs can only be found by means of an exhaustive
search among all possible patterns of a specific puncturing period
$M$. The selection of a good pattern is not intuitive, since it can
lead to catastrophic puncturing \cite{Crozier05}, i.e.,
$d^{\mathcal{C}'}_{\text{min}}\!=\!0$, or semi-catastrophic
puncturing, i.e., $z^{m}_{\text{core}}\!=\!0$ for some values of
$m$, of a constituent code $\mathcal{C}'$. Furthermore, calculation
of $d^{\mathcal{C}'}_{\text{min}}$ and
$z^{\mathcal{C}'}_{\text{min}}$ requires prior knowledge of the $M$
values of $z^{m}_{\text{core}}$.

The selection of a pseudo-random puncturing pattern guarantees that
$z^{m}_{\text{core}}\!>\!0$, and consequently,
$d^{\mathcal{C}'}_{\text{min}}\!>\!0$. Moreover,
$z^{m}_{\text{core}}$ can be expressed as a function of the memory
size $\nu$ of $\mathcal{C}'$, permitting the immediate derivation of
the minimum weights that characterize the PCCC. For the given
puncturing rate of the parity-check output, the minimum value of
$z^{m}_{\text{core}}$ is maximised, and so is
$z^{\mathcal{C}'}_{\text{min}}$, due to the properties of
pseudo-random sequences.

\begin{figure}[t]
    \centering
    \includegraphics[width=0.80\linewidth]{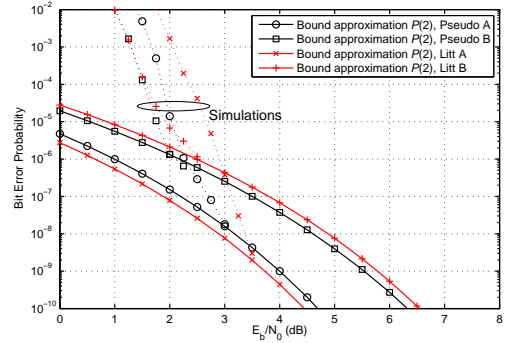}
    \caption{Bounds for various rate-1/2 PCCC(1,17/15,17/15) configurations, using an interleaver of size 1,000 bits}
    \label{BER_comparison_rate12}
\end{figure}

In Fig.\ref{BER_comparison_rate12} we have plotted the performance
of all four rate-1/2 PCCC(1, 17/15, 17/15) configurations, presented
in Table \ref{PuncPatterns}. We observe that ``Pseudo B'' slightly
outperforms the conventional ``Litt B'' configuration, while the
performance of the PCCC based on the easy to derive ``Pseudo A''
pattern is close to the performance of the PCCC based on the ``Litt
A'' pattern, obtained through exhaustive search.

\begin{table}[t]
\caption{Puncturing patterns for rate-1/2 PCCC(1,17/15,17/15)}
\label{PuncPatterns} \centering
    \begin{tabular}{|c|c|c|c|c|}
        \hline
        & Pseudo A & Pseudo B & Litt A & Litt B\\
        \hline
        Vector for Sys.Output & [1000101] & [1111101] & [0010] & [11]\\
        \hline
        Vector for 1st Par.Output & [0111010] & [0111010] & [1101] & [10]\\
        \hline
        Vector for 2nd Par.Output & [1111111] & [0111010] & [1111] & [01]\\
        \hline
    \end{tabular}
\end{table}


\section{Conclusions}
\label{conclusion}

We presented a simple approach to calculate the CWEF of punctured
RSC codes, for input sequences of minimum information weight, which
facilitates the approximation of the upper bound to the BEP, for
punctured PCCCs using large interleaver sizes. Our technique offers
the advantage of simplicity and reduced complexity, compared to
time-hungry approaches, such as brute-force, or the more complex
methods developed in \cite{Kousa02,Chatzigeorgiou06b}.

Furthermore, we considered pseudo-random puncturing patterns as a
case study for our technique and we demonstrated that they prevent
catastrophic or semi-catastrophic puncturing and facilitate the
calculation of the minimum output weights of a turbo encoder, which
characterize the performance of PCCCs. We concluded that
pseudo-random puncturing could be used to obtain rate-1/2 PCCCs
exhibiting low error floors, while specific puncturing patterns that
achieve either a lower error floor or quicker convergence to the ML
performance bound, could be determined by a subsequent search.

\bibliographystyle{IEEEtran}
\bibliography{IEEEabrv,TurboRef}

\begin{thebibliography}{10}
\providecommand{\url}[1]{#1}
\csname url@rmstyle\endcsname
\providecommand{\newblock}{\relax}
\providecommand{\bibinfo}[2]{#2}
\providecommand\BIBentrySTDinterwordspacing{\spaceskip=0pt\relax}
\providecommand\BIBentryALTinterwordstretchfactor{4}
\providecommand\BIBentryALTinterwordspacing{\spaceskip=\fontdimen2\font plus
\BIBentryALTinterwordstretchfactor\fontdimen3\font minus
  \fontdimen4\font\relax}
\providecommand\BIBforeignlanguage[2]{{%
\expandafter\ifx\csname l@#1\endcsname\relax
\typeout{** WARNING: IEEEtran.bst: No hyphenation pattern has been}%
\typeout{** loaded for the language `#1'. Using the pattern for}%
\typeout{** the default language instead.}%
\else
\language=\csname l@#1\endcsname
\fi
#2}}
\renewcommand\BIBentryALTinterwordstretchfactor{4}

\bibitem{Berrou96}
C.~Berrou and A.~Glavieux, ``Near optimum error correcting coding and decoding:
  Turbo codes,'' \emph{{IEEE} Trans. Commun.}, vol.~44, pp. 1261--1271, Oct.
  1996.

\bibitem{Divsalar95}
D.~Divsalar, S.~Dolinar, R.~J. McEliece, and F.~Pollara, ``Transfer function
  bounds on the performance of turbo codes,'' JPL, Cal. Tech., TDA Progr. Rep.
  42-121, Aug. 1995.

\bibitem{Benedetto96a}
S.~Benedetto and G.~Montorsi, ``Unveiling turbo codes: Some results on parallel
  concatenated coding schemes,'' \emph{{IEEE} Trans. Inform. Theory}, vol.~42,
  pp. 409--429, Mar. 1996.

\bibitem{Benedetto96b}
------, ``Design of parallel concatenated convolutional codes,'' \emph{{IEEE}
  Trans. Commun.}, vol.~44, no.~5, pp. 591--600, May 1996.

\bibitem{Acikel99}
{\"{O}}.~A{\c{c}}ikel and W.~E. Ryan, ``Punctured turbo-codes for {BPSK/QPSK}
  channels,'' \emph{{IEEE} Trans. Commun.}, vol.~47, pp. 1315--1323, Sept.
  1999.

\bibitem{Babich02}
F.~Babich, G.~Montorsi, and F.~Vatta, ``Design of rate-compatible punctured
  turbo ({RCPT}) codes,'' in \emph{Proc. {ICC}'02}, N.Y., USA, Apr. 2002, pp.
  1701--1705.

\bibitem{Kousa02}
M.~A. Kousa and A.~H. Mugaibel, ``Puncturing effects on turbo codes,''
  \emph{Proc. {IEE} Comm.}, vol. 149, pp. 132--138, June 2002.

\bibitem{FanMo99}
M.~Fan, S.~C. Kwatra, and K.~Junghwan, ``Analysis of puncturing pattern for
  high rate turbo codes,'' in \emph{Proc. {MILCOM}'99}, USA, Oct. 1999, pp.
  547--500.

\bibitem{Land00}
I.~Land and P.~Hoeher, ``Partially systematic rate 1/2 turbo codes,'' in
  \emph{Proc. Int. Symp. Turbo Codes}, France, Sept. 2000, pp. 287--290.

\bibitem{Chatzigeorgiou06b}
I.~Chatzigeorgiou, M.~R.~D. Rodrigues, I.~J. Wassell, and R.~Carrasco, ``A
  novel technique for the evaluation of the transfer function of punctured
  turbo codes,'' in \emph{Proc. {ICC}'06}, Turkey, July 2006.

\bibitem{Macwilliams76}
F.~J. MacWilliams and N.~J.~A. Sloane, ``Pseudo-random sequences and arrays,''
  \emph{Proc. {IEEE}}, vol.~64, pp. 1715--1729, Dec. 1976.

\bibitem{Perez96}
L.~C. Perez, J.~Seghers, and D.~J. Costello, ``A distance spectrum
  interpretation of turbo codes,'' \emph{{IEEE} Trans. Inform. Theory},
  vol.~42, pp. 1698--1709, Nov. 1996.

\bibitem{Proakis01}
J.~G. Proakis, \emph{Digital Communications}, 4th~ed.\hskip 1em plus 0.5em
  minus 0.4em\relax McGraw-Hill, 2001.

\bibitem{Sarwate80}
D.~V. Sarwate and M.~B. Pursley, ``Crosscorrelation properties of pseudorandom
  and related sequences,'' \emph{Proc. {IEEE}}, vol.~68, pp. 593--618, May
  1980.

\bibitem{Crozier05}
S.~Crozier, P.~Guinand, and A.~Hunt, ``On designing turbo-codes with data
  puncturing,'' in \emph{Proc. Can. Workshop Inf. Theory}, Canada, 2005.

\bibitem{Blazek02}
Z.~Blazek, V.~K. Bhargava, and T.~A. Gulliver, ``Some results on partially
  systematic turbo codes,'' in \emph{Proc. {VTC}'02-Fall}, Canada, Sept. 2002,
  pp. 981--984.

\end{thebibliography}

\end{document}